\documentclass[
  reprint,
  amsmath,amssymb,
  aps,prl
]{revtex4-1}

\usepackage{graphicx} 
\graphicspath{ {figures/} } 
\usepackage{subfigure}
\usepackage{dcolumn} 
\usepackage{bm}
\usepackage[normalem]{ulem}

\usepackage[normalem]{ulem}

\usepackage{hyperref} 
\hypersetup{
    colorlinks = true,
    linkcolor = black,
    filecolor = magenta,      
    urlcolor = cyan,
    citecolor = blue,
    hypertexnames = false,
    }
    
\begin{document}
\title{Probing Jet-Medium Interactions in Heavy-Ion Collisions Using Energy-Energy Correlators}

\author{Rushil Saraswat } \thanks{These authors have equal contribution.}
\affiliation{Department of Physics and Astronomy, University of California, Los Angeles, California 90095, USA}

\author{Aditya Prasad Dash} \thanks{These authors have equal contribution.}
\affiliation{Department of Physics and Astronomy, University of California, Los Angeles, California 90095, USA}

\author{Huan Zhong Huang}
\affiliation{Department of Physics and Astronomy, University of California, Los Angeles, California 90095, USA}
\author{Gang Wang}
\affiliation{Department of Physics and Astronomy, University of California, Los Angeles, California 90095, USA}
\author{Xin-Nian Wang}
\affiliation{Key Laboratory of Quark and Lepton Physics (MOE) \& Institute of Particle Physics, Central China Normal University, Wuhan 430079, China}
\author{Zhong Yang}
\affiliation{Department of Physics and Astronomy, Vanderbilt University, Nashville, TN}

\begin{abstract}
Energy–energy correlators (EECs) provide a sensitive probe of both perturbative and nonperturbative dynamics in relativistic heavy-ion collisions. Jet–medium interactions enhance particle multiplicity within the jet cone, which must be properly accounted for when extracting the EEC of jet shower hadrons in experiments. To address this issue, we develop an augmentation method that exploits momentum conservation between the near-side and away-side regions, using $\gamma$–jet events with 0–10\% centrality in Pb+Pb collisions at \texorpdfstring{$\sqrt{s_{NN}} = 5.02$ TeV}{sNN = 5.02 TeV} simulated with the CoLBT-hydro model. This approach yields an experimentally reconstructed EEC that shows improved agreement with the EEC of hadrons originating primarily from jet parton splittings. Comparing EECs of jets from Pb+Pb and p+p collisions with different matching conditions can be sensitive to jet medium interaction dynamics, and provide a novel means to test the scenario of jet energy loss in the QGP, followed by fragmentations outside the QGP.

\end{abstract}

\maketitle


\section{\label{sec:level1}Introduction}
Heavy-ion collisions at the BNL Relativistic Heavy Ion Collider (RHIC) and the CERN Large Hadron Collider (LHC) create a deconfined state of matter known as the quark–gluon plasma (QGP), in which quarks and gluons behave as quasi-free particles~\cite{QGPViscosity2011}. Jets, collimated sprays of hadrons produced through the fragmentation and hadronization of high-momentum partons, serve as powerful probes of the QGP. As jets traverse the medium, they interact with it and deposit energy, rendering jet observables highly sensitive to the dynamical properties of the QGP~\cite{JetQuenching2016,MediumResponse2020,Kudinoor2025}.

Energy–energy correlators (EECs) have recently emerged as powerful tools for studying jet structure and jet–medium interactions in heavy-ion collisions~\cite{Neill:2022, Moult:2025}. The EEC for hadron pairs inside a jet cone is defined in terms of their two-dimensional angular separation and transverse momenta ($p_T$) as~\cite{CMS2025, ALICE2024EEC, Liang2025}
\begin{equation}
{\rm EEC}(\Delta R) \equiv \frac{1}{N_{\text{jet}}} \frac{1}{\delta r_{R}}\sum_{\substack{\text{hadrons}: \\ 2|\Delta r_{ij} - \Delta R| < \delta r_{R}}} \Big(\frac{p_{T,i} \, p_{T,j}}{p_{T, \rm jet}^2}\Big)^{n_{\rm exp}}, 
\label{eq:EEC_eqn}
\end{equation}
where the pairwise distance $\Delta r_{ij}$ between two hadrons is defined in terms of their pseudorapidity ($\eta$) and azimuthal angle ($\varphi$),
\begin{equation}
\begin{split}
\Delta r_{ij}\equiv\sqrt{(\eta_i-\eta_j)^2+(\varphi_i-\varphi_j)^2}.
\end{split}
\end{equation}
Here, $\delta r_{R}$ denotes the widths of the the bins centered at $\Delta R$. The normalization factor $N_{\rm jet}$ represents the total number of selected jets in the event sample, and $p_{T,\text{jet}}$ is the final transverse momentum of the quenched jet. In this work, we fix the energy-weighting exponent to $n_{\rm exp} = 1$. Therefore, in the following, EEC$_j$ denotes the correlator with unit energy weight unless otherwise specified.
By measuring the $p_T$-weighted angular distribution of particles within jets, the EEC directly reflects the evolution of jet substructure, from the earliest hard splittings to the formation of final-state hadrons~\cite{Dixon2022}.

Energy deposited by a propagating jet can induce a hydrodynamic wake in the medium, leading to an enhancement of soft hadrons (\(p_T \lesssim 2\)~GeV/\(c\)) along the jet direction (the wake front) and a depletion in the opposite hemisphere (the diffusion wake)~\cite{Yang2022, Tachibana2020, Park2021, Kudinoor2025}. This jet-induced medium excitation (JIME)~\cite{Chen2018} can affect EEC measurements. In particular, it may lead to differences between the EEC extracted using traditional background subtraction techniques~\cite{PHENIX2020, Hwa2004} and that of jet shower hadrons (denoted EEC$_{j,\mathrm{shower}}$), which we define as hadrons originating from the hadronization of partons produced by hard splittings of the final partons, with negligible contribution from medium excitations.
This study analyzes simulated events of 0--10\% central Pb+Pb collisions at \(\sqrt{s_{NN}} = 5.02\)~TeV generated with the CoLBT-hydro model to investigate how JIME modifies the underlying event within the jet cone and how it impacts the experimental extraction of EEC$_j$. 
We further propose a data-driven background subtraction method to account for JIME and restore EEC$_j$ to the true EEC$_{j,\mathrm{shower}}$. EEC$_{j,\mathrm{shower}}$ can be experimentally accessed to study jet evolution without contamination from the soft medium response, while also enabling the isolation of the EEC associated with the medium response inside the jet cone.

Parton energy loss in the QGP has been extensively studied at RHIC~\cite{STAR:whitepaper2005,PHENIX:whitepaper2005} and at the LHC~\cite{CMS2017, ATLAS:2018jet, ATLAS:2023gammajet, ALICE:2019jet}. The nuclear modification factor $R_{AA}$~\cite{Pop2002,Adler2003,Gyulassy2003} is commonly used to quantify medium-induced modifications of particle spectra:
\begin{equation}
R_{AA}(p_T) = \frac{dN_{AA}/dp_T}{\langle N_{\text{coll}} \rangle \, dN_{pp}/dp_T},
\end{equation}
where $\langle N_{\text{coll}} \rangle$ denotes the average number of binary nucleon--nucleon collisions. Early measurements of $R_{AA}$ focused on leading high-$p_T$ particles and were interpreted within a picture in which energetic partons are produced in initial hard scatterings, lose energy while traversing the hot and dense QGP, and subsequently hadronize outside the medium. In this framework, the fragmentation of high-$p_T$ partons occurs predominantly in the vacuum and is therefore largely decoupled from the space–time evolution of the QGP, with medium properties influencing final hadron spectra primarily through the accumulated energy loss. Extensive theoretical efforts have aimed to relate $R_{AA}$ quantitatively to characteristic features of parton energy loss~\cite{Wang:2009, Marshall:2025, Spousta:2025}, and recent empirical studies for both light and heavy quarks are broadly consistent with this picture~\cite{Marshall:2025}. In particular, these studies enable experimental constraints on how jet energy loss depends on the initial medium density and the geometric path length traversed through the QGP~\cite{Marshall:2025}.

Motivated by these findings, we further investigate the dynamics of jet fragmentation using EEC$_{j,\mathrm{shower}}$, which is sensitive to the angular and $p_T$ distributions of partons during the perturbative shower and hadronization stages~\cite{Liang2025, ALICeE2024, Yang2024, Andres2024}, while excluding the contribution from medium response due to jet--medium interaction. This observable provides complementary information to $R_{AA}$ and enables a more differential probe of the underlying parton energy-loss scenario.

The remainder of this paper is organized as follows. Section~\ref{sec:level2} introduces the CoLBT-hydro model, describes the event samples used in this study, and outlines the conventional procedure for extracting EEC$_j$. Section~\ref{sec:jime_effects} presents a method to quantify the effect of JIME on EEC$_j$ and proposes a procedure to reduce the difference between EEC$_j$ and EEC$_{j,\rm shower}$. In Sec.~\ref{sec:jet_dynamics}, we investigate $\gamma$-jet energy loss and fragmentation dynamics using the EEC as a probe. Finally, we summarize our findings in Sec.~\ref{sec:conclusion}.

\section{\label{sec:level2}Method}
\subsection{CoLBT-hydro Model}
The CoLBT-hydro model \cite{Chen2018, Chen2020, Zhao2022} is employed to simulate $\gamma$-jet propagation and the associated jet-induced medium response in Pb+Pb collisions. The framework integrates the microscopic linear Boltzmann transport (LBT) model \cite{He2015}, which describes the propagation of energetic jet shower and recoil partons, with the event-by-event (3+1)D CCNU-LBNL viscous hydrodynamic (CLVisc) model \cite{Pang2012, Pang2015, Pang2018}, which governs the evolution of the bulk medium and the soft components of the jet-induced medium excitation \cite{Yang2023}. The LBT and CLVisc components are dynamically coupled in real time through source terms that account for the energy–momentum deposited into the medium by jet shower and recoil partons, as well as backreaction effects associated with particle holes \cite{Yang2023}.

The $\gamma$-jet configurations for both Pb+Pb and $p$+$p$ collisions are generated using PYTHIA8 \cite{Sjostrand2008}. In Pb+Pb events, partons from the initial jet shower and multi-parton interactions (MPI) propagate through the QGP and induce medium response within the CoLBT-hydro framework. The final hadron spectra consist of hadrons from the hadronization of energetic partons via a parton recombination model \cite{Pang2015, Han2016, Zhao2020}, together with underlying-event hadrons obtained from the bulk medium through Cooper–Frye freeze-out.

The hydrodynamic evolution employs a freeze-out temperature $T_f = 137$ MeV and a specific shear viscosity $\eta/s = 0.15$. It uses the s95p parametrization of the lattice QCD equation of state with a rapid crossover \cite{Huovinen2010}, together with Trento initial conditions \cite{Moreland2015} supplemented by a longitudinal envelope at an initial time $\tau_0 = 0.6$ fm/$c$ \cite{Yang2023}. These parameters are tuned to reproduce experimental data on bulk hadron spectra and anisotropic flow observables at LHC energies \cite{Pang2018, Yang2023}.
Pb+Pb and $p$+$p$ collision events at $\sqrt{s_{NN}} = 5.02$ TeV are simulated with $\gamma$-jet selection cuts $|\eta_{\rm jet}| < 1.6$ and $|\eta_{\gamma}| < 1.44$, and the jet radius parameter is set to $R_j = 0.5$.

\subsection{Event Samples}
\label{sec:samples}

In this study, for each Pb+Pb event containing a $\gamma$-jet (referred to as the signal event, $S$), a corresponding hydrodynamic event (denoted $HS$) is generated using identical initial conditions but without embedding the $\gamma$-jet. The jet cone in $S$ contains two types of charged particles: jet shower hadrons ($h_j^{\mathrm{ch}}$) \cite{Hwa2004} and underlying soft charged hadrons ($h_U^{\mathrm{ch}}$). The latter primarily originate from the bulk medium and are largely uncorrelated with the hard probe; these constitute the underlying background ($US$). However, $h_U^{\mathrm{ch}}$ also includes a smaller contribution from medium response correlated with the $\gamma$-jet.

A sample is constructed of Pb+Pb events without a $\gamma$-jet at the same \texorpdfstring{$\sqrt{s_{NN}}=5.02$ TeV}{sNN=5.02 TeV}.  The events $M_1$, $M_2$, and $M_U$ are selected from this sample within the same centrality class as the $\gamma$-jet signal event and are required to have a charged-particle multiplicity within 0.1\% of that of $S$ in a kinematic region $P$. The region $P$ is defined by $|\Delta\eta| > 0.5$ and $0\leq|\Delta\varphi| < 2\pi$, 
where $\Delta \eta \equiv \eta-\eta_{j,{\rm Pb}}$, $\Delta \varphi \equiv \varphi-\varphi_{j,\rm{Pb}}$, and $(\eta_{j,\mathrm{Pb}}, \phi_{j,\mathrm{Pb}})$ denote the jet-axis coordinates in the corresponding Pb+Pb event. The events $M_{1,2,U}$ are rotated around the beam axis to match the reaction-plane orientation of the signal event, thereby closely reproducing the underlying azimuthal structure of $S$ in the absence of JIME.

A Pb+Pb event sample with an embedded $p$+$p$ collision is also constructed to simulate heavy-ion collisions containing a jet without medium-modified substructure or JIME. To obtain the jet shower hadrons for this hybrid event, a $p$+$p$ event with a $\gamma$-jet is selected such that its $p_{T,\gamma}$ matches that of S within 5\%. The coordinates of hadrons inside the jet cone in the selected $p$+$p$ event, $(\eta_{pp}, \varphi_{pp})$, are shifted to align with the jet axis of the corresponding Pb+Pb event according to
\begin{eqnarray}
 \eta_{pp}^{\rm embed} &\equiv&  \eta_{pp}+\eta_{j,{\rm Pb}}-\eta_{j,pp}, \\
\varphi_{pp}^{\rm embed} &\equiv& \varphi_{pp}+\varphi_{j,{\rm Pb}}-\varphi_{j,pp}, 
\end{eqnarray}
where $(\eta_{j,pp}, \phi_{j,pp})$ denote the jet-axis coordinates in the selected $p$+$p$ event. These shifted hadrons are then superimposed onto $M_U$, thereby generating a heavy-ion event with an embedded $p$+$p$ $\gamma$-jet process but without jet-induced medium excitation.


\subsection{EEC of charged jet shower hadrons}

For CoLBT events, we can obtain the jet shower hadrons within the jet cone and apply Eq.~(\ref{eq:EEC_eqn}) to directly calculate EEC$_{j,\mathrm{shower}}$. This is the EEC of hadrons originating predominantly from the hadronization of partons produced via hard splittings of final partons.

In experimental measurements, the EEC is constructed from unique pairings of all charged hadrons within the jet cone, as jet shower hadrons cannot be unambiguously separated from particles originating from the underlying event. The resulting pair combinations can be decomposed as
\begin{equation}
\begin{split}
S \times S=h_j^{\rm ch}\times h_j^{\rm ch}+h_U^{\rm ch}\times h_U^{\rm ch}+h_U^{\rm ch}\times h_j^{\rm ch},
\end{split}
\label{eq:eectot}
\end{equation}
where $h_j^{\rm ch}\times h_j^{\rm ch}$ represents the genuine jet shower contribution, while $h_U^{\rm ch}\times h_U^{\rm ch}$ and $h_U^{\rm ch}\times h_j^{\rm ch}$ correspond to combinatorial background contributions arising from underlying-event particles and mixed pairs of underlying and jet shower particles, respectively.

To isolate the jet shower component ($h_j^{\rm ch}\times h_j^{\rm ch}$) from the total $S\times S$, the traditional procedure introduced in Ref.~\cite{CMS2025} is employed. We refer to this as the  unaugmented background subtraction method, defined as
\begin{equation}\label{eq:eec_extraction}
\begin{split}
{\rm EEC}_j= S \times S- S\times M_1- M_1\times M_1+ M_1\times M_2 .
\end{split}
\end{equation}
where $A \times B$ denotes the EEC calculated from unique pairings of hadrons taken from events $A$ and $B$ within the jet cone defined by $\Delta\eta^2+\Delta\varphi^2 \leq R_j^2$. Here, $M_1$ and $M_2$ represent MB events without any $\gamma$-jet. The same $M_1$ and $M_2$ selected for $S$ are also used when calculating EEC$_j$ in the corresponding $p$+$p$ embedded events.

\begin{figure}
\centering
\includegraphics[width=1\linewidth]{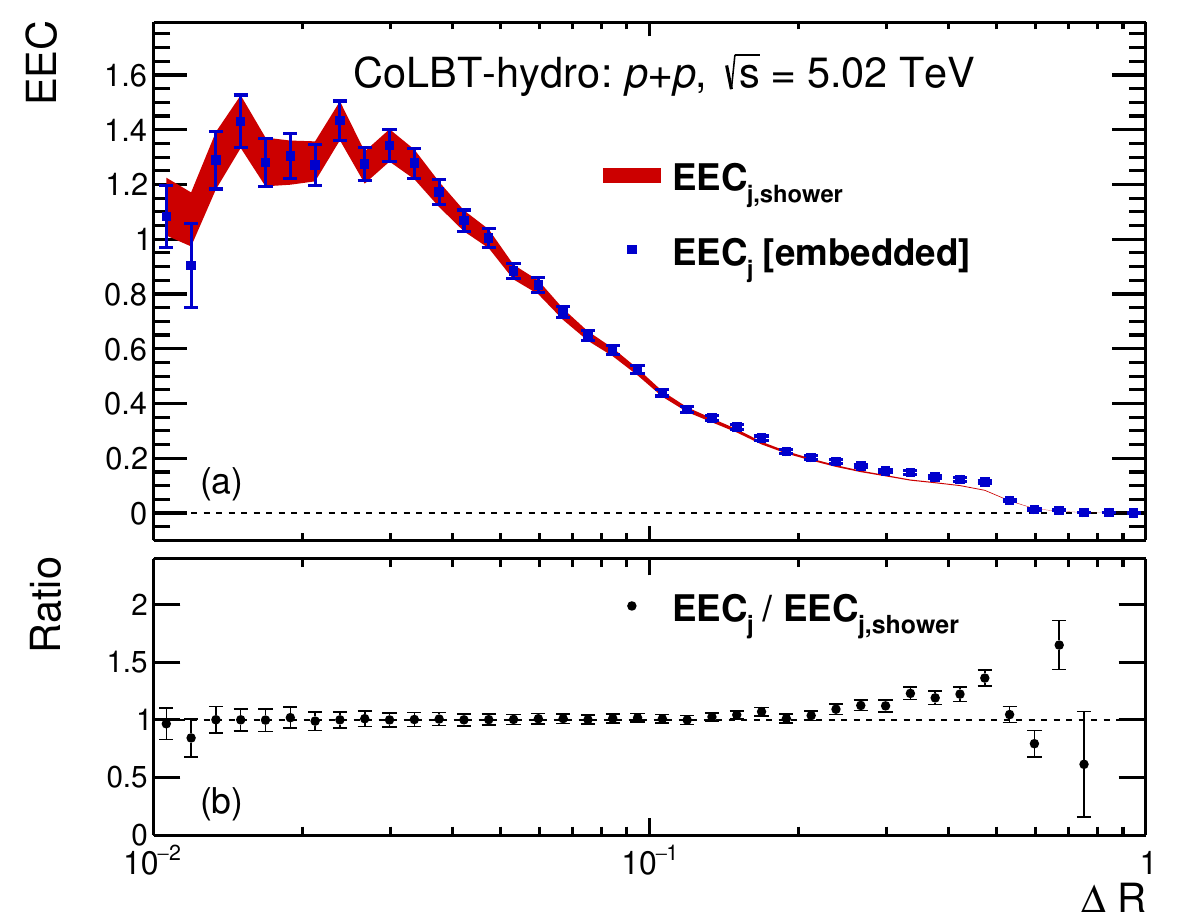}
\caption{CoLBT-hydro simulations of (a) EEC$_{j,{\rm shower}}$ of charged jet shower hadrons in the jet cone from $p$+$p$ collisions at $\sqrt{s}=5.02$ TeV (red band) and EEC$_j$ extracted using the unaugmented background subtraction method from 0-10\% central Pb+Pb collisions at $\sqrt{s_{NN}}=5.02$ TeV embedded with $p$+$p$ $\gamma$-jet events (blue solid squares) and (b) their ratio (black solid circles).}
    \label{fig:eecfullrange}
\end{figure}

Figure~\ref{fig:eecfullrange} compares the CoLBT-hydro calculations of EEC$_{j,\mathrm{shower}}$ of charged jet shower hadrons in the jet cone from $p$+$p$ collisions at $\sqrt{s}=5.02$ TeV (red band) with EEC$_j$ extracted using the unaugmented background subtraction method from $0$–$10\%$ central Pb+Pb collisions at $\sqrt{s_{NN}}=5.02$ TeV embedded with $p$+$p$ $\gamma$-jet events (blue solid squares). The close agreement between the two demonstrates that, in the absence of JIME, the unaugmented background subtraction method reliably reproduces EEC$_{j,\mathrm{shower}}$.

However, the ${\rm EEC}_j$ obtained from Eq.~(\ref{eq:eec_extraction}) generally contains contributions not only from jet shower hadrons but also from hadrons originating from medium-scattered partons associated with JIME. Consequently, an additional procedure is required to isolate the pure jet shower correlation EEC$_{j,\mathrm{shower}}$ from ${\rm EEC}_j$. In the following analysis, we focus on the region $\Delta R > 0.1$ dominated by perturbative processes to further explore the effect of medium interaction with jet partons on EEC$_j$ \cite{ALICE2024EECp}.

\section{\label{sec:jime_effects}Effect of JIME on EEC$_j$}

The wake front associated with JIME has been extensively studied experimentally at the LHC~\cite{CMSjime,ATLAS:2024diffusion}. This wake can lead to an enhancement of soft hadrons within the jet cone in heavy-ion collisions~\cite{He2021, He2020}. Such an enhancement is present in the signal ($S$) event sample but absent in the MB samples $M_{1,2}$. As a result, the background subtraction may underestimate the underlying contribution inside the jet cone, causing EEC$_j$ to exhibit an excess relative to EEC$_{j,\rm shower}$ in Pb+Pb collisions.

\begin{figure}
\centering
\includegraphics[width=1\linewidth]{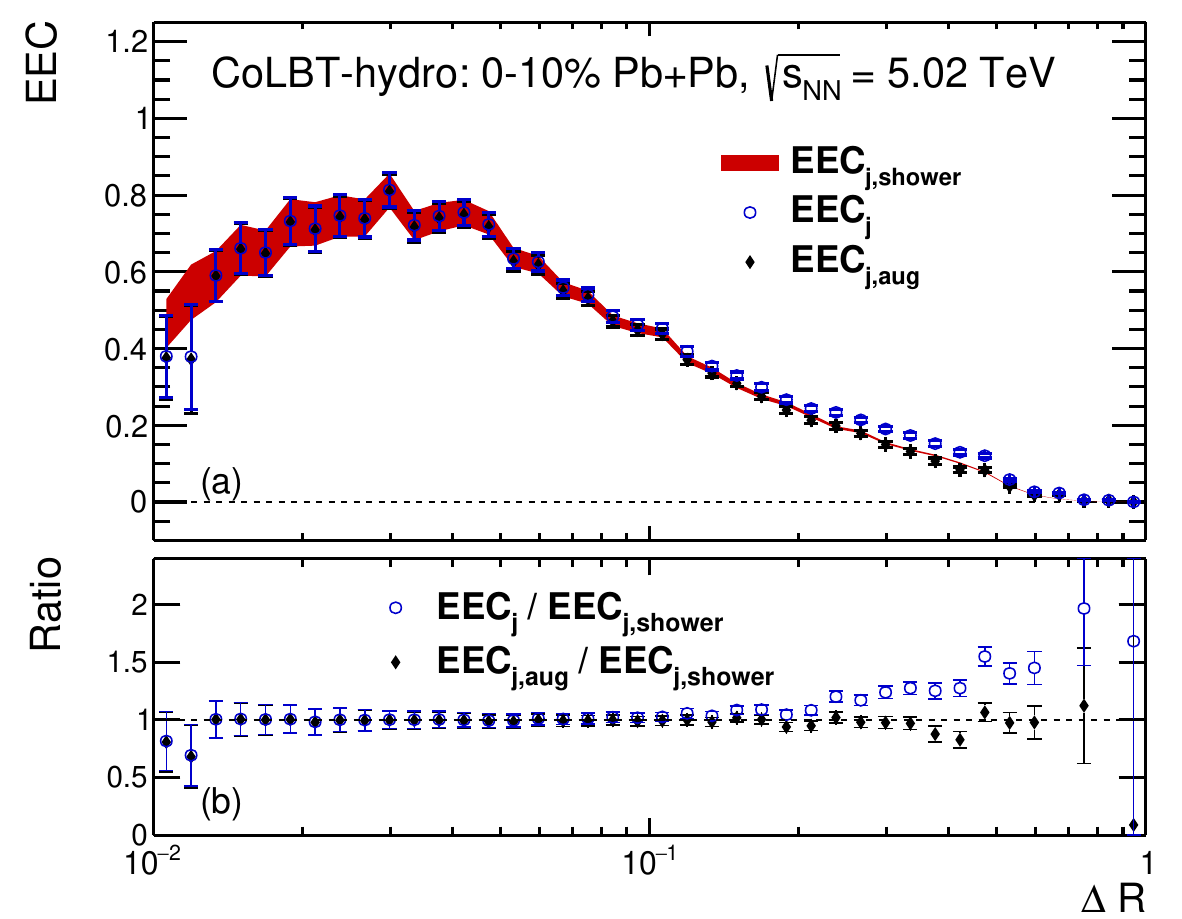}
\caption{CoLBT-hydro simulations of (a) EEC$_{j,{\rm shower}}$ of charged jet shower hadrons in the jet cone from 0--10\% centrality Pb+Pb events at $\sqrt{s_{NN}}=5.02$ TeV (red band), EEC$_j$ extracted using the unaugmented background subtraction method (blue open circles), and EEC$_{j,{\rm aug,\ [model]}}$ using the model-augmented background subtraction method (black solid diamonds) and (b) the ratios of EEC$_{j}$ (blue open circles) and EEC$_{j,{\rm aug}}$ (black solid diamonds) with EEC$_{j,{\rm shower}}$.}
\label{fig:modelaugeec}
\end{figure}

Figure~\ref{fig:modelaugeec} compares the CoLBT-hydro simulations of the true EEC$_{j,\rm shower}$ in $0$–$10\%$ central Pb+Pb collisions (red band) with EEC$_j$ extracted using the unaugmented background subtraction method (blue hollow circles).
For the latter, the combinatorial background is estimated using MB Pb+Pb events embedded with corresponding $p$+$p$ $\gamma$-jet events, $M_{1,2,U}$, as described in Sec.~\ref{sec:samples}.
The extracted EEC$_j$ is significantly larger than EEC$_{j,{\rm shower}}$ in the perturbative region. This excess is attributed to multiplicity modifications induced by jet showers and JIME, while simulation with no embedded jet ($M_U$) underestimates the combinatorial background inside the jet cone.



To account for the multiplicity difference between the HS and US samples (see Sec.~\ref{sec:samples}), we add a weight factor $w_{i,j}$ in the definition of EEC and modify Eq.~(\ref{eq:EEC_eqn}) as follows
\begin{equation}
{\rm EEC}(\Delta R) \equiv \frac{1}{N_{\text{jet}}} \frac{1}{\delta r_{R}}\sum_{\substack{\text{hadrons}: \\ 2|\Delta r_{ij} - \Delta R| < \delta r_{R}}} w_{i,j}\Big(\frac{p_{T,i} p_{T,j}}{p_{T, \rm jet}^2}\Big)^{n_{\rm exp}}, 
\label{eq:EEC_eqn_weight}
\end{equation}
where $w_{i,j} \equiv A(h_i)A(h_j)$. The augmentation factor $A(h)$ is constructed from the ratio of charged-hadron yields ($Y_{h}$) between the full $US$ and $HS$ event samples in each $(\Delta\eta,\Delta\phi)$ bin~\cite{Yang2023},
\begin{equation}
\begin{split}
A(h)=\begin{cases}
Y^{US}_{h}/Y^{HS}_{h}, & h \in M_{1,2} \\
1, & h \notin M_{1,2} 
\end{cases}.
\end{split}
\label{eq:afact}
\end{equation}
We refer to this approach as the model-augmented background subtraction method, and denote the resulting observable as EEC$_{j,{\rm aug ,\ [model]}}$. Figure~\ref{fig:modelaugeec} shows that EEC$_{j,{\rm aug}}$ (black solid diamonds) is significantly closer to EEC$_{j,\rm shower}$ than the unaugmented result, EEC$_j$.


However, in experimental analyses, $HS$ and $US$ event samples cannot be reliably identified. To overcome this limitation, we develop a procedure to construct an $HS$ sample using the $M_U$ sample as a proxy, referred to as the data-augmented background subtraction method. This method exploits the multiplicity depletion in the opposite hemisphere (away-side) to estimate the enhancement inside the jet cone (near-side).
We make three assumptions:
(\textbf{I}) The signal sample $S$ provides a good approximation of the $US$ sample on the away side, where jet shower hadrons are negligible~\cite{Hanks:2012}. (\textbf{II}) The $US$ sample consists predominantly of soft hadrons. (\textbf{III}) The near-side enhancement of soft-hadron $p_T$ is balanced by a corresponding depletion in an away-side cone defined by $\Delta\eta^2+(\Delta\varphi-\pi)^2 \leq R_{\rm away}^2$. 
$R_{\rm away}$ is chosen such that the vector sum of the $\vec{p}_T$ enhancement of soft charged hadrons from the near- and away-side cones, defined as the difference between hadrons in $S$ and those in $M_U$, balances the $\vec{p}_T$ lost by the jet ($p_{T,\rm jet} - p_{T,\gamma}$). We further restrict to events in which the component of this vector sum along the $\gamma$ axis is less than $10$ GeV/$c$. Here, ``soft" refers to $p_T < 4$ GeV/$c$.

Under these assumptions, the charged-hadron yield ratio ($A(h)$) in a given $(\Delta\eta,\Delta\varphi)$ bin on the near side can be estimated from the soft-hadron yields in several $(\Delta\eta_{\rm away},\Delta\varphi_{\rm away})$ bins on the away side. To establish this correspondence, the away-side cone is shifted by $\pi$ in azimuth so that it is centered on the near-side jet axis, and then uniformly rescaled by a factor $R_j/R_{\rm away}$ to match the near-side cone radius. The $(\Delta\eta_{\rm away},\Delta\varphi_{\rm away})$ bins are scaled and mapped accordingly. After this geometric transformation, the near-side bin receives contributions from the away-side bins that overlap with it. The yield ratio can thus be rewritten as
\begin{align}
\frac{Y^{US}_{h}}{Y^{M_U}_{h}}
&= 1+ \frac{Y^{US}_{h}-Y^{M_U}_{h}}
           {Y^{M_U}_{h}}  \nonumber \\
&\approx  1+ \sum_{(\Delta\eta_{\rm away},\Delta\varphi_{\rm away})}
\frac{Y^{M_U}_{h,\rm soft}
      -Y^{US}_{h,\rm soft}}
     {Y^{M_U}_{h,\rm soft}}  \nonumber \\
&\approx  1+ \sum_{(\Delta\eta_{\rm away},\Delta\varphi_{\rm away})}
\left(
1-\frac{Y^{S}_{h,\rm soft}}
        {Y^{M_U}_{h,\rm soft}}
\right),
\end{align}
where the summation runs over the mapped away-side bins associated with the given near-side $(\Delta\eta,\Delta\varphi)$ bin.

\begin{figure}
\centering
\includegraphics[width=1\linewidth]{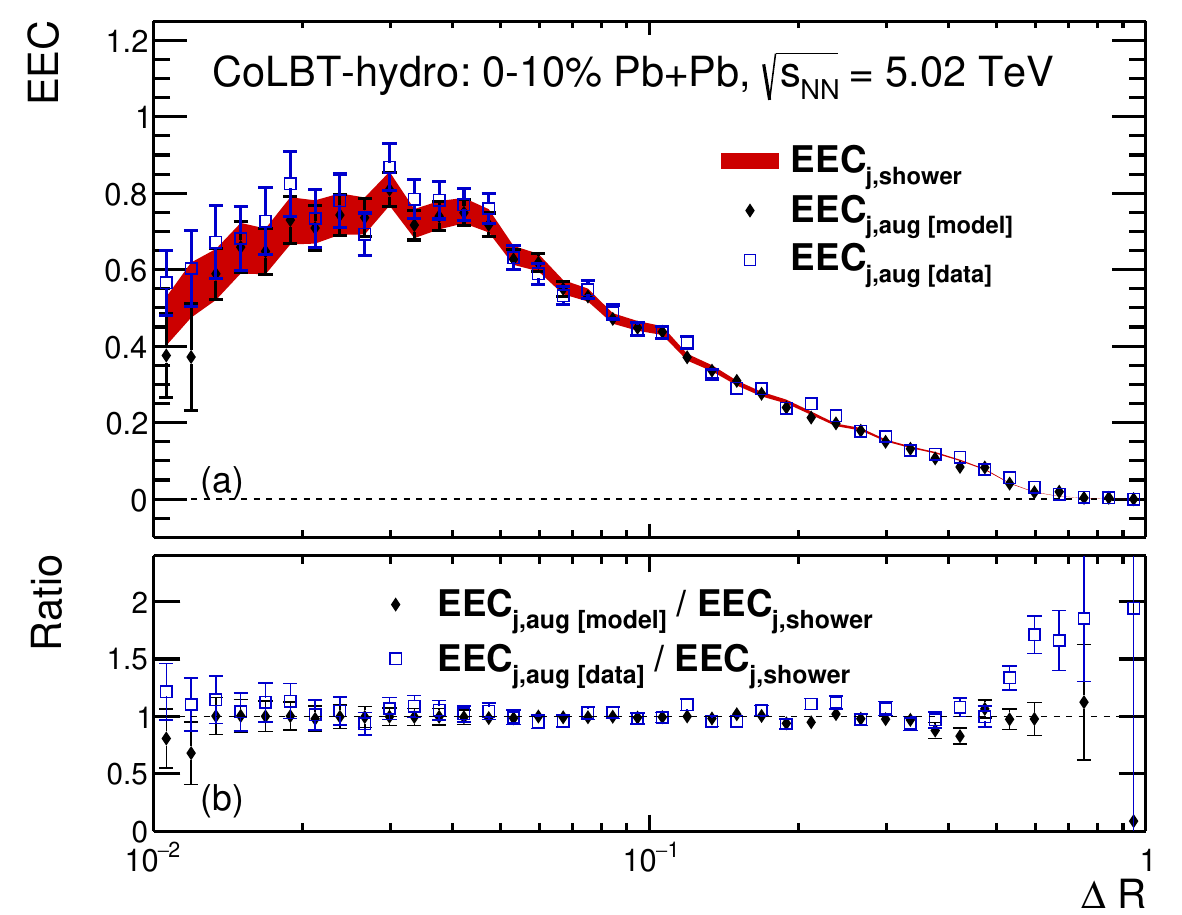}
\caption{CoLBT-hydro simulations of (a) EEC$_{j,{\rm shower}}$ of charged jet shower hadrons in the jet cone from 0--10\% centrality Pb+Pb events at $\sqrt{s_{NN}}=5.02$ TeV (red band), EEC$_{j,{\rm aug}}$ using the model-augmented background subtraction method (black solid diamonds), and EEC$_{j,{\rm aug}}$ using the data-augmented background subtraction method (blue open squares) and (b) the ratios of EEC$_{j,{\rm aug\, [model]}}$ (black solid diamonds) and EEC$_{j,{\rm aug\, [data]}}$ (blue open squares) with EEC$_{j,{\rm shower}}$}.
    \label{fig:augeec}
\end{figure}

Figure~\ref{fig:augeec} is similar to Fig.~\ref{fig:modelaugeec}, except that the EEC$_j$ obtained with the unaugmented background subtraction method is removed and replaced by EEC$_{j,{\rm aug,\ [data]}}$ extracted using the data-augmented background subtraction method (blue hollow squares).
The data-augmented method, like the model-augmented approach, reproduces the expected jet shower EEC from the CoLBT simulations with good accuracy within the jet cone. This suggests that the proposed data-augmented background subtraction method can be applied experimentally to extract the EEC of jet shower hadrons in the presence of JIME.
Applying this method to different theoretical models in future studies will allow a systematic evaluation of its model dependence.


We note that jet–medium interactions inevitably mix partons originating from the initial hard-scattered parton with those from the hot medium. In this work, we adopt the CoLBT definition of jet shower particles for the EEC calculation within the jet cone.
We observe noticeable differences in the perturbative region arising from JIME and from the various background subtraction methods. In the remainder of this paper, we further investigate the sensitivity of the EEC in the perturbative region to the underlying parton evolution dynamics. 

\section{\label{sec:jet_dynamics}EEC and Dynamics of Parton Energy Loss in the QGP}

We can employ the data-augmented background subtraction method described in Section~\ref{sec:jime_effects} to approximate EEC$_{j,\rm shower}$. Jet–medium interactions, including parton energy loss in the QGP and subsequent hadronization, are modeled using CoLBT-hydro-simulated $\gamma$-jet events in 0--10\% central Pb+Pb collisions at $\sqrt{s_{NN}}=5.02$~TeV. For comparison, we also simulate $\gamma$-jet events in $p$+$p$ collisions $\sqrt{s}=5.02$~TeV, which provide a baseline for jet evolution in the absence of medium-induced energy loss. By focusing on the features of EEC$_{j,\rm shower}$ in the perturbative region, we investigate how jet hadronization is modified in a dynamical scenario that includes parton energy loss.

\begin{figure*}[bthp]
 \centering
\includegraphics[width=1\linewidth]{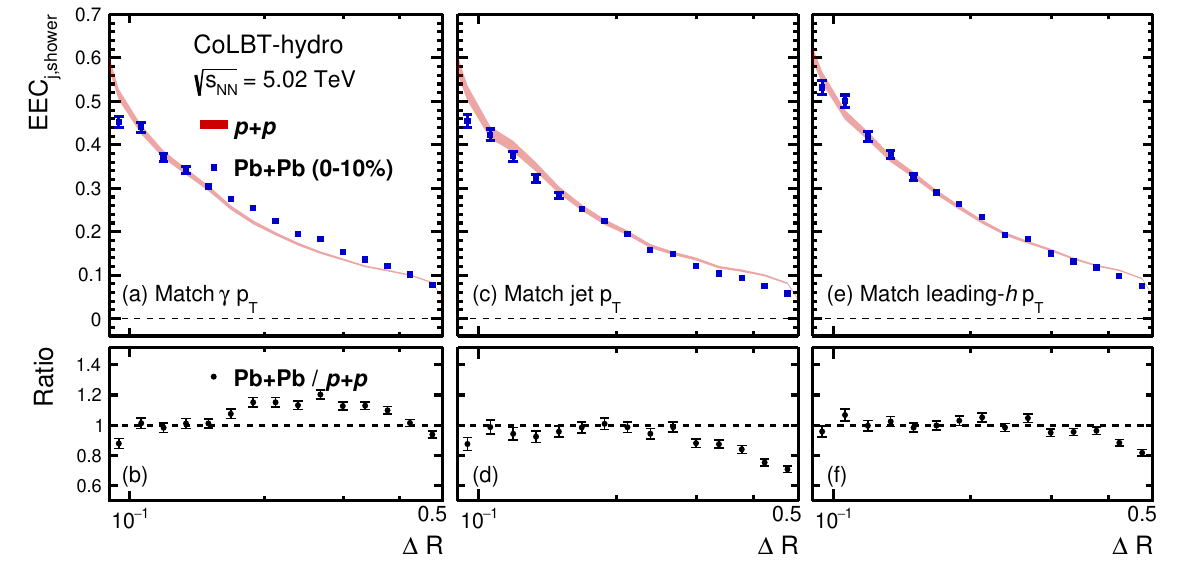}
\caption{CoLBT-hydro simulations of EEC$_{j,\rm shower}$ of charged jet shower hadrons in the perturbative region for 0--10\% central Pb+Pb at $\sqrt{s_{NN}}=5.02$ TeV and $p$+$p$ collisions at $\sqrt{s}=5.02$ TeV in panels (a), (c), and (e) and their ratios in panels (b), (d), and (f) respectively. Panel pairs \{a, b\} correspond to the case where the photon $p_T$ in $p$+$p$ events is required to match, within 5\%, that in Pb+Pb events, \{c, d\} correspond to the case where the photon $p_T$ in $p$+$p$ events is required to match, within 5\%, the quenched jet $p_T$ in Pb+Pb events after energy loss, and \{e, f\} correspond to the case where the leading-hadron $p_T$ in $p$+$p$ events is required to match, within 5\%, that in Pb+Pb events.}
\label{fig:eec_verif}
\end{figure*}




Figure~\ref{fig:eec_verif}(a) first compares EEC$_{j,\rm shower}$ for $\gamma$-jet processes in the perturbative region for central Pb+Pb and $p$+$p$ collisions from CoLBT simulations. The initial photon $p_T$ in Pb+Pb events is required to match that in $p$+$p$ events within 5\%, thereby ensuring comparable initial jet $p_T$ in the two systems. Under this condition, a significant difference is observed in the EEC$_{j,\rm shower}$ constructed from final-state, evident when the ratio in Fig.~\ref{fig:eec_verif}(b) deviates from unity. This difference may reflect medium-induced modifications of the jet fragmentation pattern due to parton energy loss in the QGP. However, it could also arise, at least in part, from the mismatch between the quenched jet $p_T$ in Pb+Pb and the corresponding jet $p_T$ in $p$+$p$, since energy loss shifts the final jet energy and thus alters the effective comparison basis.

To reduce the impact of the quenched jet-$p_T$ mismatch, Fig.~\ref{fig:eec_verif}(c) shows EEC$_{j,\rm shower}$ with a different selection criterion: the $\gamma$ $p_T$ in $p$+$p$ collisions is required to match, within 5\%, the quenched jet $p_T$ after energy loss in Pb+Pb collisions. In this case, the quenched jet in Pb+Pb has approximately the same final jet $p_T$ as the jet in the $p$+$p$ event. The overall difference in EEC$_{j,\rm shower}$ between Pb+Pb and $p$+$p$, indicated by the ratio in Fig.~\ref{fig:eec_verif}(d), is reduced compared to the initial-$p_T$ matching case, although a slightly enhanced deviation remains around $\Delta R \sim 0.5$.

Figure~\ref{fig:eec_verif}(e) applies a more differential matching condition, requiring that the $p_T$ of the leading hadron in the $p$+$p$ jet matches, within 5\%, that of the leading hadron in the Pb+Pb jet. With this leading-hadron matching criterion, the selected $\gamma$-jet samples from Pb+Pb and $p$+$p$ exhibit improved consistency of EEC$_{j,\rm shower}$ distributions, demonstrated by reduced deviation from unity in the ratio as shown in Fig.~\ref{fig:eec_verif}(f) in comparison with Fig.~\ref{fig:eec_verif}(b,d). We restrict to the region $\Delta R < 0.5$ in Fig.~\ref{fig:eec_verif}, as edge effects can significantly distort the shape of EEC$_{j,\mathrm{shower}}$ at larger $\Delta R$, where the cross section is very small~\cite{Andres:2512.10026}.
 
Taken together, these results are consistent with a dynamical picture in which partons lose energy while traversing the QGP medium, whereas the hadronization of high-$p_T$ partons occurs predominantly outside the medium in the CoLBT framework. The leading-hadron $p_T$ matching appears to provide a more robust correspondence between Pb+Pb and $p$+$p$ events than matching based on reconstructed quenched jet $p_T$, possibly reflecting practical uncertainties in jet $p_T$ determination. Overall, the comparison of EEC$_{j,\rm shower}$ between heavy-ion and $p$+$p$ collisions offers a sensitive probe of the dynamical mechanisms of parton energy loss in the QGP, and its application to experimental data could help test the hypothetical dynamical picture and provide valuable insight into jet–medium interactions.

\section{\label{sec:conclusion}Conclusion}

We presented various methods of measuring jet energy-energy correlators (EEC$_j$) in heavy-ion collisions and evaluated their sensitivity to JIME and parton energy loss in the QGP. Using $\gamma$-jet events simulated with the CoLBT-hydro model, we systematically studied how medium response and jet quenching affect the extraction and interpretation of EEC observables.

We first demonstrated that traditional background subtraction methods for EEC$_j$ in Pb+Pb collisions can be biased by JIME. The hydrodynamic wake induced by jet energy deposition enhances the soft hadron multiplicity inside the jet cone and modifies the underlying event, leading to deviations between the measured EEC$_j$ and the intrinsic correlator of jet shower hadrons (EEC$_{j,\rm shower}$). To mitigate this effect, we proposed a data-augmented background-subtraction method that re-weights the minimum-bias multiplicity distribution to account for the JIME-induced enhancement. This approach significantly improves the extraction of the jet-shower EEC and brings the reconstructed EEC$_{j,{\rm aug}}$ into close agreement with EEC$_{j,\rm shower}$ from the model. Our study suggests that this method can be implemented in experimental analyses to better isolate jet fragmentation from medium response effects.

We then explored the dynamical implications of EEC$_{j,\rm shower}$ by comparing Pb+Pb and $p$+$p$ events under different matching conditions. When events are selected with similar initial photon $p_T$, a significant difference in EEC$_{j,\rm shower}$ is observed, which can originate from both genuine medium modification of the fragmentation pattern and the mismatch of the quenched jet $p_T$ after energy loss. Instead, by matching the final jet $p_T$, the difference in the perturbative region of the EEC is reduced. Furthermore, when the leading-hadron $p_T$ is matched between Pb+Pb and $p$+$p$ events, the EEC$_{j,\rm shower}$ distributions become nearly identical. Our simulation results are consistent with the dynamical picture in which energetic partons lose energy while traversing the QGP, and the hadronization processes of the quenched high-$p_T$ partons occur predominantly outside the medium. 

Overall, our study demonstrates that jet energy-energy correlators provide a sensitive probe of both jet-induced medium excitation and the space-time dynamics of parton energy loss. With proper treatment of background and medium response, EEC measurements in heavy-ion collisions can offer valuable insight into the interplay between perturbative parton shower evolution and the strongly coupled QGP medium. Future studies extending this analysis to inclusive jet samples, different centrality classes, and varying collision energies at RHIC and the LHC will further elucidate the microscopic mechanisms of jet–medium interactions.

\section{\label{sec:5} Acknowledgements}

Aditya Prasad Dash, Huan Zhong Huang and Gang Wang are supported by the U.S. Department of Energy under Grant No. DE-FG02-88ER40424. Xin-Nian Wang is supported by NSFC under Grant No. 12535010. The simulation data from the CoLBT-hydro model were generated at NSC3/CCNU.

\bibliographystyle{unsrt}
\bibliography{ref}
\end{document}